\documentclass[runningheads]{svmult}
 
\usepackage{makeidx}   
\usepackage{graphicx}  
\usepackage{subeqnar}  
\usepackage{multicol}  
\usepackage{physprbb}  
\makeindex             
 
\begin{document}
\title*{The Globular Cluster Luminosity Function}
\toctitle{The Globular Cluster Luminosity Function}
%
%
\titlerunning{The Globular Cluster Luminosity Function}
%
\author{Dean E.~McLaughlin}
\authorrunning{Dean E.~McLaughlin}
%
%
\institute{Space Telescope Science Institute, 3700 San Martin Drive,
Baltimore, MD 21218 USA}
 
\maketitle              
 
\begin{abstract}
The main aspects of the globular cluster luminosity function needing to
be
explained by a general theory of cluster formation are reviewed, and the
importance of simultaneously understanding globular cluster systematics
(the fundamental plane) within such a theory is pointed out.
\end{abstract}

\section{Review}

A clear
understanding of the physics driving the basic form of the globular cluster
luminosity function, or GCLF -- the distribution of cluster magnitudes,
luminosities, or masses in a galaxy -- remains elusive. To be sure,
substantial progress has been made in
the theory of globular cluster mass loss and dynamical evolution over a
Hubble time in galaxies; in our ideas about the assembly of large galaxies
from multiple smaller fragments; in our understanding of how dense
pockets of gas are converted in general to stars and star clusters; and
in our appreciation of the origin and evolution of self-gravitating
structure in turbulent gas. But the GCLF ultimately is shaped to some extent
by every one of these processes and relies on a complex interplay between
them.

\begin{figure}[t]
\begin{center}
\vspace{-1.5truein}
\includegraphics[width=0.9\textwidth]{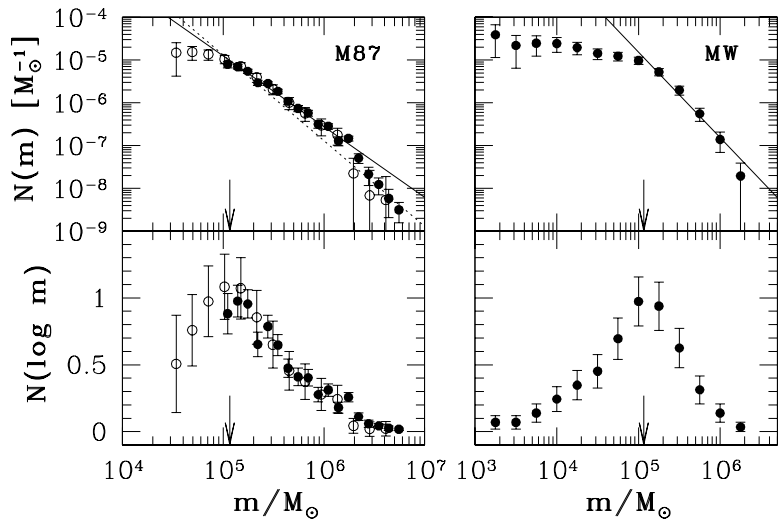}
\end{center}
\caption[]{Observed mass spectra and GCLFs of M87 and the Milky Way. Absolute
cluster magnitudes have been converted to masses using a constant
$M/L_V=1.45\ M_\odot\,L_\odot^{-1}$ \cite{mcl00a}. Data for M87 are taken
from \cite{mhh94}
(filled points, coming from $\simeq$0.8--4.5 effective radii in the galaxy)
and \cite{hhm98} (open points, from 0.1--1.5 effective radii). Those for the
Milky Way are taken from \cite{har96}. Normalizations are chosen so as
to put exactly one cluster at the ``turnover'' mass,
$m_*\simeq1.2\times10^5\,M_\odot$, marked by arrows}
\label{fig1}
\end{figure}

Traditionally, the GCLF was constructed by plotting the number of globulars
around a galaxy in equal-sized bins of integrated cluster magnitude. Since
the globulars in our own Galaxy, at least, are known to share a common core
mass-to-light ratio \cite{mcl00a}, the result is equivalent to the mass
distribution $N(\log\,m)\equiv dN/d\,\log\,m$. This is the function with
the familiar, Gaussian-like appearance, shown in the lower panels of
Fig.~\ref{fig1} for the globular cluster systems of M87 and the Milky
Way. The location of its peak, at $m_*\simeq1.2\times10^5\,M_\odot$,
corresponds to the classic GCLF ``turnover'' magnitude $M_V^0=-7.4$, which
serves remarkably well as a standard candle (see, e.g., the review in
\cite{har01}) and was interpreted in the first serious theories of globular
cluster formation \cite{pd68} \cite{fr85} as the imprint of a Jeans mass
set by specific thermal conditions at pre- or protogalactic epochs.
However, when clusters are counted instead in intervals of equal
\emph{linear} mass, the physical distribution obtained is $N(m) \equiv
dN/dm =m^{-1}N(\log\,m)$, which is necessarily different in shape from the
usual GCLF. This function, generally referred
to as the globular cluster \emph{mass spectrum}, is shown for M87 and
the Milky Way in the top panels of Fig.~\ref{fig1}. There it is clear that,
while $m_*$ retains some physical significance as a point of sharp change in
the physical
dependence of $N(m)$, the number of globulars per unit mass continues to
increase down to the lowest observed masses. It is then no longer obvious that
there exists any ``preferred'' globular cluster mass scale \cite{hp94}.
 
Disagreement persists among theoretical studies (e.g., compare
\cite{fz01} with \cite{ves98} \cite{ves00} \cite{ves01}) as to whether
the change of slope in $N(m)$ around $m=m_*$ had to have been
established almost immediately at the time of cluster formation,
or if instead $N(m)$ might initially have risen towards $m<m_*$ just as
steeply as it falls towards $m>m_*$, with its current form resulting
entirely from evaporation and tidal shocks removing globulars from the
low-mass end of the distribution. But however this issue resolves itself,
the same discussions support the idea that the
shape of $N(m)$ above $m>m_*$ reflects reasonably well
the initial distribution. Theories of globular cluster formation should
therefore aim to address three main points taken from Fig.~\ref{fig1} 
(see also \cite{hp94} \cite{mp96a}):
 
{\bf (1)} Over about a decade in globular
cluster mass above $m>m_*$, $N(m)$ can be approximated
by power laws with exponents in the range $-1.7$ to $-2$ [$N(m)\propto
m^{-1.65}$ is drawn as a solid line in the upper M87 panel of
Fig.~\ref{fig1}, and $N(m)\propto m^{-2}$ in the upper Milky Way panel],
which are strikingly similar to those describing the
mass distributions of giant
molecular clouds (GMCs) in the Milky Way, of the dense star- and
cluster-forming ``clumps'' inside GMCs, and of the super star
clusters currently forming in mergers such as the Antennae (where
$N_{\rm SSC}\propto m_{\rm SSC}^{-2}$ \cite{zf99}). Thus, any theory of the
GCLF ought to be a special case of a more general theory of structure and
star (cluster) formation in the interstellar medium.
 
{\bf (2)} While $m_*$ is essentially identical in
M87 and the Milky Way (as in many other galaxies), the
overall shape of the globular cluster mass spectrum is \emph{not}
universal: The $m^{-2}$ power law that applies in the Milky Way and the
Antennae (and M31: \cite{mp96a}) is also shown (as a dotted line) against
the observed $N(m)$ in M87. It clearly fails to describe those data, and
the same is true in several other giant ellipticals \cite{hp94}.
It is difficult to ascribe this difference to the effects of dynamical
evolution, since much of the M87 data are taken from beyond one effective
radius in the galaxy (where the GCLF should be closer to its original form
than in the inner regions \cite{ves00}), and since GCLF evolution is
expected to have been less drastic in larger galaxies in the first place
\cite{mw97} \cite{ves00}. Thus, any theory for the origin of the GCLF must
allow for real differences in the shape of $N(m)$ from galaxy to galaxy but
leave little room for variations in the mass $m_*$.
 
{\bf (3)} There is no detectable variation in either the value of $m_*$
or the shape of $N(m)$ above $m_*$ as a function of galactocentric
position inside M87 \cite{mhh94} \cite{hhm98} \cite{kun99}.
Thus, a theory of globular cluster formation should couple
with a realistic scenario of galaxy formation and evolution to produce a
GCLF that is fairly insensitive to intragalactic environment.
 
Recent studies of dynamical evolution of the GCLF \cite{ves98}
\cite{ves00} \cite{ves01} \cite{fz01} appear (modulo some disagreements in
detail, and only by assuming a host galaxy potential that is
static over a Hubble time) to have made promising headway in
understanding the third of these items. However, such calculations always
assume the initial form of the GCLF \emph{a priori}; they do not aim to
address either of points (1) or (2). Two models that do have been developed
in the literature.

Elmegreen and collaborators construct an essentially geometric theory
(no underlying dynamics are drawn upon or constrained) relating the
fractal dimension of structure in any turbulent interstellar medium to a
power-law mass spectrum for star- and cluster-forming gas clouds, thus
exploiting
similarities between our local
interstellar medium and the GCLF \cite{elm97} \cite{elm96}. This approach
suffered at first from an oversimplified view of the fractal geometry of both
turbulence and the real interstellar medium (see, e.g., \cite{cha01}), as well
as an overly strict expectation of universality in the globular cluster mass
spectrum [cf.~point (2) above]. However, recent refinements \cite{elm02}
stand potentially to remove these concerns.
 
Harris \& Pudritz \cite{hp94} and McLaughlin \& Pudritz \cite{mp96a} instead
build on older models of structure in the local interstellar medium and
calculate the steady-state spectrum of gaseous protoglobular cluster (PGC)
masses that develops through a competition between mass build-up by
coalescent PGC-PGC collisions vs.~the destruction of massive PGCs
by feedback from their own star formation. The shape of $N(m)$ is
then determined primarily by two fitting parameters: a mass dependence in the
feedback destruction timescale ($\tau_d$) of the larger PGCs and the
ratio ($\beta$) of a fiducial self-destruction time to a typical PGC
collision timescale. Good fits to the observed $N(m)$ above $m_*$ in the
globular cluster systems of both M87 and the Milky Way can be obtained
\cite{mp96a} \cite{har01}, but the required behaviour of $\tau_d$ and the
values of $\beta$ are strictly \emph{ad hoc}.

Both of these models really concern themselves with a description of the
mass spectrum of gaseous protoclusters and assume that the globulars
themselves will directly inherit the PGC $N(m)$. Thus, they require
that the star-formation efficiency ($\epsilon \equiv m_{\rm GC}/m_{\rm PGC}$)
in any PGC be \emph{independent of its initial mass}. They also connect
explicitly with descriptions of the local interstellar medium and current
star formation by applying scalings between physical parameters (masses,
radii, and velocity dispersions) for turbulent PGCs that are \emph{virialized
and in pressure equilibrium with an ambient medium}. Just how well justified
these assumptions are, is a question that arises from the consideration of
other properties of globular clusters.

\section{The Globular Cluster Fundamental Plane}

\begin{figure}[b]
\begin{center}
\vspace{-1.65truein}
\includegraphics[width=0.9\textwidth]{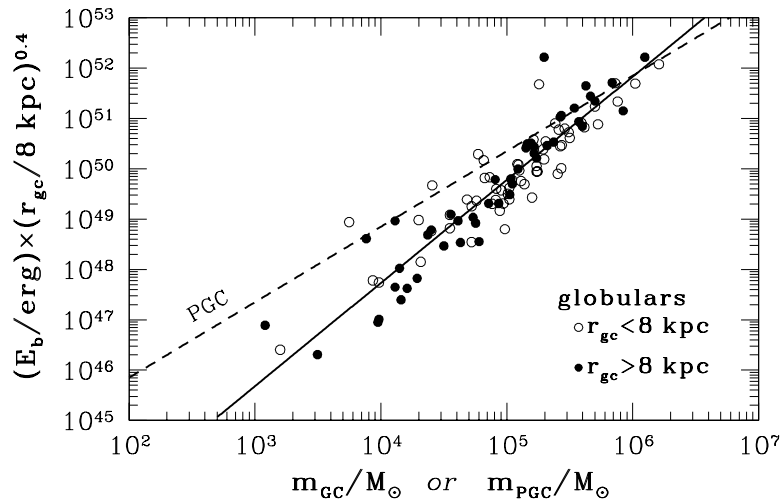}
\end{center}
\caption[]{Binding energy vs.~mass as observed for Milky Way globular
clusters \cite{mcl00a} and expected for virialized protoglobulars
under high ambient pressure}
\label{fig2}
\end{figure}

The multiple correlations between structural parameters of Milky Way globular
clusters can be shown to reduce to just two independent relations that define
a fundamental plane \cite{djo95} and that physically signify \cite{mcl00a}
a constant mass-to-light ratio in the cluster cores  ($M/L_V=1.45\
M_\odot\,L_\odot^{-1}$) and a tight scaling between cluster binding
energy, total luminosity or mass, and Galactocentric position:
\begin{equation}
E_b({\rm GC}) = 3.4\times10^{39}\ {\rm erg}\
\left(m_{\rm GC}/M_\odot\right)^{2.05}
\left(r_{\rm gc}/8\,{\rm kpc}\right)^{-0.4}\ \ ,
\label{eq1}
\end{equation}
which is drawn through the Galactic globular cluster data (normalized to a
single Galactocentric radius of 8 kpc) in Fig.~\ref{fig2}.
Globular clusters in M31, M33, and NGC 5128 (Cen A) appear to fall along
the binding energy-mass relation defined by the Galactic
system, and in fact extend it to higher cluster masses \cite{bar02}
\cite{lar02} \cite{hhhm02}.
But if gaseous protoglobulars were -- like the star-forming clumps in
Galactic GMCs today -- in virial equilibrium within a pressurized
ambient medium but strongly self-gravitating (i.e., at the maximum
equilibrium mass allowed under a given surface
pressure $P_s$), then their binding energies should have shown quite a
different dependence on total mass; and this ultimately suggests that the
globular cluster $N(m)$ might have differed significantly from the original
protocluster distribution even before any dynamical evolution set in.

From the general relations given in \cite{mp96b}, PGCs of the type
just described would have obeyed the relation
\begin{equation}
E_b({\rm PGC}) \simeq 7.2\times10^{42}\ {\rm erg}\ 
\left(m_{\rm PGC}/M_\odot\right)^{3/2} 
\left(P_s/10^8\,k_{B}\right)^{1/4}\ \ .
\label{eq2}
\end{equation}
The pressure scale $P_s=10^8\ k_B$ is
significantly higher than typical pressures in our local
interstellar medium, but this has been suggested as a necessary condition for
the formation of bound, massive stellar clusters (e.g., \cite{elm97}
\cite{az01}) and so is worth explicit consideration. The
dashed line in Fig.~\ref{fig2} therefore traces equation (\ref{eq2})
with $P_s$ fixed, as a representative PGC energy-mass relation.
(Changing $P_s$ even by as much as two orders of magnitude would move
the line only slightly.)

If stars form in direct proportion to local gas density, then
$E_b\propto m^2/R$ implies that $E_b({\rm GC})/E_b({\rm PGC}) \propto
\epsilon^2 \left(R_{\rm PGC}/R_{\rm GC}\right)$ for $\epsilon$
the star-formation efficiency in a PGC. The ratio of cluster to protocluster
radii is further related to $\epsilon$, in a way that depends on details of
the massive-star feedback that ends star formation in a PGC and clears it of
any remaining gas \cite{hil80} \cite{ad00}. Under the high ambient pressure
$P_s$ specified here, the freefall time of any PGC will be $\sim10^6$ yr
or less \cite{az01}, suggesting that feedback might operate on a timescale
longer than the dynamical time. In this case \cite{hil80}, $R_{\rm PGC}/
R_{\rm GC}=\epsilon$, and so $E_b({\rm GC})/E_b({\rm PGC}) \propto
\epsilon^3$. Then,
noting that $P_s\propto r_{\rm gc}^{-2}$ if the ambient gas around the
Galactic PGCs was distributed in an isothermal potential well, the ratio
of equations (\ref{eq1}) and (\ref{eq2}) is nearly independent of
Galactocentric position: $\epsilon^3 \propto E_b({\rm GC})/E_b({\rm PGC})
\propto m_{\rm GC}^{2.05}/m_{\rm PGC}^{1.5}\propto \epsilon^{2.05}
m_{\rm PGC}^{0.55}$, and thus $\epsilon\equiv \left(m_{\rm GC}/
m_{\rm PGC}\right)\propto m_{\rm PGC}^{0.58}$. This scaling obviously has to
saturate at some protocluster mass for which star formation is 100\%
efficient. The value of this mass is not known from first principles (nor
has any mass dependence in $\epsilon$ been suggested from first
principles), but in this highly simplified example it seems reasonable to
assert that $\epsilon=1$ where the energy-mass relations of
the globulars and PGCs intersect, i.e., around $m\sim 10^6\,M_\odot$.

A variation of star-formation efficiency with protocluster gas mass was
suggested by \cite{mcl00b} as the origin of the steeper $E_b - m$ slope
for globulars relative to PGCs, and it was discussed in considerably more
detail by \cite{az01}, as the potential cause of the different mass-radius
relations among globulars (roughly, $R\propto m^0$ \cite{mcl00a}) and
virialized gas clouds ($R\propto m^{1/2}$ \cite{mp96b}). These two approaches
are clearly equivalent, and the results of \cite{az01} are identical to
those found here. But to return to the main point, if any of this is even
roughly correct, and $\epsilon$ increases with PGC mass over most of the
mass range of the observed GCLF, then the PGC mass spectrum, $N(m_{\rm PGC})$,
had to have been steeper than the globular cluster $N(m_{\rm GC})$ obtained
immediately after star formation and feedback were completed. The details of
this are given in \cite{az01}, from which it can be seen that if
$N(m_{\rm GC})\propto m_{\rm GC}^{-\alpha}$ and $\epsilon\propto
m_{\rm PGC}^{0.6}$, then $N(m_{\rm PGC})\propto m_{\rm PGC}^{-\beta}$ with
$\beta=1.6\alpha-0.6$ (for $m_{\rm PGC}<10^6\,M_\odot$ or so). Thus,
for $\alpha=1.65$, as is found above $m_{\rm GC}>m_*$ in the M87 globular
cluster system, $N(m_{\rm PGC})\propto m_{\rm PGC}^{-2}$. But in the Milky 
Way (and M31, and the Antennae), the unevolved parts of the globular cluster
spectrum already follow $\alpha=2$, and $N(m_{\rm PGC})\propto
m_{\rm PGC}^{-2.6}$ is implied. This slope is significantly steeper
than any found or theorized for structures in the interstellar medium
as we know it, and it presents a
clear problem if we wish to hold on to the idea that globular cluster
formation was simply a special case of generic star formation in a familiar
interstellar medium.

It is possible that the removal of gas from a PGC by stellar feedback is not
``slow'' compared to the star-formation or dynamical times (e.g., if the
pressure on protocluster surfaces were not as high as $10^8\ k_B$, these
internal timescales would not be so short). But even though $\epsilon$ would
not necessarily depend as strongly on $m_{\rm PGC}$ in such a case, the
qualitative problem would remain of $N(m_{\rm PGC})$ being steeper than
$N(m_{\rm GC})$, unless some other aspect of the analysis
is also in error. Moreover, if the dynamical timescale of a protocluster
grows to $\sim10^7$ yr or more, there may be time for massive stars to
explode as supernovae and dispel most of the gas before appreciable amounts
of it can be converted to stars. This could make it difficult to produce
tightly bound star clusters at all. (The energy injected by a single
supernova into surrounding gas is $\sim10^{51}$ erg, and one Type II
supernova is expected for every $135\,M_\odot$ of stars formed with a
Salpeter initial mass function. The combined energy exceeds the binding
energy of any PGC in Fig.~\ref{fig2} by orders of magnitude.)

It may simply be that the inferred increase of $\epsilon$ with $m_{\rm PGC}$
is the result of an incorrect assumption that the star formation
inside a protocluster traces the local gas density. Perhaps
gas clouds of all masses form stars in the same overall proportion, but
more massive ones produce them in more centrally concentrated configurations.
If so, then the GCLF above $m_*$ could indeed accurately reflect the PGC
mass spectrum (as current models assume), which would in turn be explained
within conventional theories of the interstellar medium. The binding
energy--mass relation for globulars (not to mention the putative constancy
of $\epsilon$) would still have to be produced from first principles in a
theory of star formation with rigorous feedback calculations.

It could also be that the presumed scaling of PGC binding energy
with mass, $E_b({\rm PGC})\propto m_{\rm PGC}^{3/2}$, is incorrect because
the dense gas clumps that produce bound stellar clusters do not originate in
a state of virial or pressure equilibrium. There are a few interrelated
points suggesting that this may be the case: the high pressures
($P_s\sim 10^8\ k_B$) that are sometimes invoked \cite{elm97} \cite{az01}
to allow for various aspects of globular cluster formation are in excess
of values calculated for any equilibrium setting; the problem of supernova
energetics mentioned just above makes it almost inevitable that globulars
(or any similarly tightly bound stellar cluster) had to have formed very
rapidly from the collapse of the highest peaks in a field of density
fluctuations, and these could plausibly have decoupled from pressure
equilibrium with the ambient background; and in the most
modern view of the interstellar medium as a large-scale turbulent flow or
cascade (the sort of backdrop against which \cite{elm96} \cite{elm97} and
\cite{elm02} work to compute mass spectra for protoclusters), hydrostatic
equilibrium in transient substructures seems less than guaranteed. 
But if PGCs were not in virial or pressure equilibrium, then it is not
obvious that they couldn't themselves have followed something like
$E_b({\rm PGC})\propto m_{\rm PGC}^2$ or so -- in which case the 
observed $E_b - m$ scaling now seen in globulars presents no difficulties.

Whatever the true explanation of Fig.~\ref{fig2}, it seems clear that
further advances in our understanding of the globular cluster luminosity
function will come only alongside a better appreciation of the process of
star formation at a more detailed level. It is essential that attention
be paid simultaneously to the multiple empirical clues provided by various
properties of globular cluster systems.

%
 
\end{document}